\documentclass [a4paper,fleqn]{article}

\usepackage {amsmath} \usepackage{amssymb} \usepackage{cite} \usepackage{amsthm}
\numberwithin{equation}{section} \numberwithin{table}{section} \mathindent=0pt
\theoremstyle{plain} \newtheorem{theorem}{Theorem}

\numberwithin{theorem}{section}

\begin{document}

\title{Nonlinear differential equations with exact \\ solutions expressed via the Weierstrass function}
\author{N.A. Kudryashov}
\date{Department of Applied Mathematics\\
Moscow  Engineering and Physics Institute\\
(State university)\\
31 Kashirskoe Shosse,  115409\\
Moscow, Russian Federation} \maketitle

\begin{abstract}
New problem is studied that is to find nonlinear differential
equations with special solutions expressed via the Weierstrass
function. Method is discussed to construct nonlinear ordinary
differential equations with exact solutions. Main step of our
method is the assumption that nonlinear differential equations
have exact solutions which are general solution of the simplest
integrable equation. We use the Weierstrass elliptic equation as a
building block to find a number of nonlinear differential
equations with exact solutions. Nonlinear differential equations
of the second, third and fourth order with special solutions
expressed via the Weierstrass function are given. Most of these
equations are used at the description of nonlinear waves in
physics.
\end{abstract}



\section{Introduction}
Last years we can observe a splash of papers with methods of
finding exact solutions of nonlinear differential equations
\cite{1,2,3,4,5,6,7,8,9}. There are two reasons to make the study
of this direction. First there is a great interest to the
investigation of nonlinear processes. Secondly we have codes
\emph{Maple, Mathematica} and other ones to have a lot of
symbolical calculations.

It is well known that all nonlinear differential equations can be
conditionally divided into three types: exactly solvable,
partially solvable and those that have no exact solution.

Usually investigators solve two problems in the theory of
nonlinear differential equations. First problem is to find
nonlinear solvable differential equations and to study their
properties. The second problem is to search exact solutions of
nonlinear differential equations.

This paper is devoted to solution of new problem that is to find
nonlinear differential equations with special solutions. It is
important to note that these equations have exact solutions but
they are not all integrable equations. Using our approach we
extend class of studied differential equations.

Consider nonlinear evolution equation

\begin{equation}
\label{1.1}E_1[u] \equiv E_1(u, u_t, u_x, ..., x, t)=0
\end{equation}

Assume we need to have exact solutions of this equation. Usually
we look for exact solution of nonlinear evolution equation taking
 the travelling wave into account and search exact solution of
equation \eqref{1.1} in the form

\begin{equation}
\label{1.2}u(x, t) = y(z), \qquad z=x-C_0 t
\end{equation}

As a result we have that the equation \eqref{1.1} reduces to the nonlinear ordinary
differential equation (ODE)

\begin{equation}
\label{1.3}E_2[y] \equiv E_2(y, y_z, ...,z)=0
\end{equation}

To obtain exact solutions of equation \eqref{1.3} one can apply
different approaches \cite{10,11,12,13,14,15,16,17}. However one
can note that the most methods that are used to search exact
solutions take the singular analysis into account  for solutions
of the nonlinear differential equations.

Using the singular analysis first of all one can consider the
leading members of equation \eqref{1.3}. After that one can find
the singularity for solution of equation \eqref{1.3}. Further the
truncated expansion is used to have the transformation to search
exact solutions of nonlinear ODEs. At this point one can use some
trial functions (hyperbolic and elliptic and so on) to look for
exact solutions of nonlinear ODEs.

However one can note that hyperbolic and elliptic functions are
general solutions of nonlinear exactly solvable equations. We have
as a rule that partially solvable nonlinear differential equations
have exact solutions that are general solutions of solvable
equations of lesser order.

This paper is the extension of our recent work \cite{18} where we
started the solution of the new problem that is to find nonlinear
ordinary differential equations of polynomial form which have
special solutions expressed via general solution of the Riccati
equation.

Let us explain the idea of this work. It is well known there is the great problem that is
to find nonlinear integrable differential equations. It is very important because we
often want to have general solution of nonlinear ordinary equations. However sometimes we
can content ourselves with some special solutions because a lot of differential equations
are nonintegrable ones although they are intensively used in physics and look as simple
equations. It is important to find some special solutions of these equations that are
called exact solutions. In this paper we want to find nonlinear differential equations
that are not all integrable but have special solutions in the form of the Weierstrass
function.

The first aim of this work is to present the method to find
nonlinear differential equations with exact solutions in the form
of the Weierstrass function. The second aim of this paper is to
give nonlinear ordinary differential equations that have exact
solutions expressed via general solution of the Weierstrass
elliptic equation.

The outline of this paper is as follows. In section 2 we present
the method to find nonlinear ordinary differential equations
(ODEs) with special solutions expressed via the Weierstrass
function. Nonlinear ODEs with exact solutions of the first,
second, third and fourth degree singularities are given in section
3, 4, 5 and 6. Example of nonlinear ODE with exact solution of the
fifth degree singularity that is popular at the description of the
model chaos is considered in section 7.

\section{Method applied}

Let us discuss the method that can be applied to find nonlinear
differential equations with exact solutions expressed via the
Weierstrass function. One can note that most nonlinear ordinary
differential equations has exact solutions that are general
solutions of differential equations of lesser order. Much more
than that exact equations for the most nonlinear differential
equations are determined via general solution of the Riccati
equation. This is so indeed because one can note that the most
approaches to search exact solutions of nonlinear ordinary
differential equations are based on general solution of the
Riccati equation. The application of the tanh method confirms this
idea \cite{18,19,20,21}. We gave these nonlinear differential
equations in our recent work \cite{22}.

However one can note that a lot of nonlinear ODEs have exact
solutions that are general solutions of the Weierstrass elliptic
function \cite{1,12,14,15,23}. In this connection in this paper we
are going to find nonlinear ODEs with exact solutions expressed
via the Weierstrass function.

The Weierstrass elliptic equation can be presented in the form

\begin{equation}
\label{2.1}P_1[R]=R_z^2 -4 R^3 + g_2 R+g_3 =0
\end{equation}

We have the following simple theorem.

\begin{theorem}
\label{T:2.1.} Let $R(z)$ be solution of equation \eqref{2.1} than
equations

\begin{equation}
\label{2.2}R_{zz} =6 R^2 - \frac 12 g_2
\end{equation}

\begin{equation}
\label{2.3}R_{zzz} =12RR_z
\end{equation}

\begin{equation}
\label{2.4}R_{zzzz} =120R^3 -18g_2R -12g_3
\end{equation}

\begin{equation}
\label{2.5}R_{zzzzz} =360 R^2 R_z-18 g_2 R_z
\end{equation}
have special solutions expressed via general solution of equation
\eqref{2.1}.
\end{theorem}

\begin{proof}Theorem 2.1 is proved by differentiation of \eqref{2.1} with respect to $z$
and substitution $R^2_z$ from equation \eqref{2.1} and so on into expressions obtained.

\end{proof}

It is well known that the general solution of equation \eqref{2.1}
is the Weierstrass function

\begin{equation}
\label{2.13}R(z) = \wp(z,g_{{2}},g_{{3}})
\end{equation}
where $g_{{2}}$ and $g_{{3}}$ are arbitrary constants that are
called invariants.

We want to find nonlinear ordinary differential equation which
have special solutions that are determined via general solution of
the Weierstrass elliptic equation.

Algorithm of our method can be presented by \emph{four steps.} At
\emph{the first step} we choose the singularity of special
solution and give the form of this solution. At \emph{the second
step} we take the order of nonlinear ordinary differential
equation what we want to search. \emph{The third step} lies in the
fact that we write the general form of nonlinear differential
equation taking the singularity of the solution into account and
the given order for nonlinear differential equation. \emph{The
fourth step} contains calculations. As a result we find
limitations for the parameters in order for nonlinear differential
equation has exact solutions. At this step we can have nonlinear
ODE with exact solutions.

\section{Nonlinear ODEs with exact solutions of the first degree singularity}

Let us demonstrate our approach to find nonlinear ODEs with exact
solutions of the first degree singularity. These solutions can be
presented in the form

\begin{equation}
\begin{gathered}
\label{3.31}y(z)=A_{{0}}+A_{{1}}{\frac {R_{{z}}}{R}}
\end{gathered}
\end{equation}

Here $R(z)$ is a solution of the Weierstrass elliptic equation,
$A_{{0}}$ and $A_{{1}}$ are unknown constants. First of all we are
going to find nonlinear second order ODEs.

\emph{Second order ODEs.} General form of the second order ODEs
can be presented as the following

\begin{equation}
\begin{gathered}
\label{3.32}y_{{{\it
zz}}}+a_{{1}}yy_{{z}}+a_{{2}}{y}^{3}+b_{{0}}y_{{z}}+
b_{{1}}{y}^{2}-C_{{0}}y+C_{{1}}=0
\end{gathered}
\end{equation}

Equation \eqref{3.32} was written taking singularity of solution
\eqref{3.31} into account and the given order of nonlinear ODE
what we want to have with exact solutions.

Let us assume $A_{{1}}=1$. Substituting $y(z)$ from \eqref{3.31} into \eqref{3.32}  and
using \eqref{2.1} and \eqref{2.2} we have equation in the form

\begin{equation}
\begin{gathered}
\label{3.33}\left( 2\,b_{{0}}+12\,a_{{2}}A_{{0}}+4\,b_{{1}}+2\,a_{{1}}A_{{0}}
 \right) {R}^{4}+\\
 \\
+ \left( 4\,a_{{2}}R_{{z}}+2\,R_{{z}}+a_{{2}}{A_{{0}}}
^{3}-C_{{0}}A_{{0}}+2\,a_{{1}}R_{{z}}+C_{{1}}+b_{{1}}{A_{{0}}}^{2}
 \right) {R}^{3}+\\
 \\
 + \left( 2\,b_{{1}}A_{{0}}R_{{z}}-3\,a_{{2}}A_{{0}}g_{
{2}}-b_{{1}}g_{{2}}+3\,a_{{2}}{A_{{0}}}^{2}R_{{z}}-C_{{0}}R_{{z}}
 \right) {R}^{2}+\\
 \\
 + \left( b_{{0}}g_{{3}}+a_{{1}}A_{{0}}g_{{3}}-a_{{2}}R
_{{z}}g_{{2}}-3\,a_{{2}}A_{{0}}g_{{3}}-b_{{1}}g_{{3}} \right) R-\\
\\
-a_{{2} }R_{{z}}g_{{3}}-2\,R_{{z}}g_{{3}}+a_{{1}}R_{{z}}g_{{3}}=0
\end{gathered}
\end{equation}

From the last equation we get relations for constants

\begin{equation}
\begin{gathered}
\label{3.34}a_{{1}}=-1-2\,a_{{2}},\quad
C_{{0}}=3\,a_{{2}}{A_{{0}}}^{2}+2\,b_{{1}}A_{{0}}, \quad g_{{2}}=0
\end{gathered}
\end{equation}

\begin{equation}
\begin{gathered}
\label{3.37}a_{{2}}=-1, \quad A_{{0}}=\frac15 b_{{0}}+ \frac25 b_{{1}}
\end{gathered}
\end{equation}

\begin{equation}
\begin{gathered}
\label{3.39}C_{{1}}=-{\frac {7}{125}}\,{b_{{0}}}^{2}b_{{1}}-{\frac {4}{125}}\,b_{{0
}}{b_{{1}}}^{2}+{\frac {4}{125}}\,{b_{{1}}}^{3}-{\frac {2}{125}}\,{b_{ {0}}}^{3}, \quad
b_{{1}}=-3\,b_{{0}}
\end{gathered}
\end{equation}

Taking these constants into account we have equation in the form

\begin{equation}
\begin{gathered}
\label{3.41}y_{{{\it zz}}}+yy_{{z}}-{y}^{3}+b_{{0}}y_{{z}}-3\,b_{{0}}{y}^{2}-3\,{b
_{{0}}}^{2}y-{b_{{0}}}^{3}=0
\end{gathered}
\end{equation}

 Solution of this equation is expressed by the formula

\begin{equation}
\begin{gathered}
\label{3.42}y \left( z \right) =-b_{{0}}+{\frac {R_{{z}}}{R}}
\end{gathered}
\end{equation}

At $b_{{0}}=0$ from \eqref{3.41} we have equation that was found by P. Painleve
\cite{24}.

\emph{Third order ODEs. }The general case of the third order ODEs
with exact solutions  \eqref{3.31} takes the form

\begin{equation}
\begin{gathered}
\label{3.43}a_{{0}}y_{{{\it zzz}}}+a_{{1}}yy_{{{\it zz}}}+a_{{2}}{y_{{z}}}^{2}+a_{
{3}}{y}^{4}+b_{{0}}y_{{{\it zz}}}+b_{{1}}yy_{{z}}+b_{{2}}{y}^{3}+\\
\\
+d_{{0 }}y_{{z}}+d_{{1}}{y}^{2}-C_{{0}}y+C_{{1}}=0
\end{gathered}
\end{equation}

Assuming $A_{{1}}=1$ without loss of the generality and
substituting \eqref{3.31} into \eqref{3.43} we have coefficients
in the form

\begin{equation}
\begin{gathered}
\label{3.45}b_{{2}}=-\frac12 a_{{1}}A_{{0}}-\frac12 b_{{1}}-\frac12
b_{{0}}-4\,a_{{3}}A_{{0 }}
\end{gathered}
\end{equation}

\begin{equation}
\begin{gathered}
\label{3.46}C_{{0}}=-\frac 32 b_{{1}}{A_{{0}}}^{2}+2\,d_{{1}}A_{{0}}-8\,a_{{3}}{A_{{0}
}}^{3}-\frac32{A_{{0}}}^{3}a_{{1}}-\frac32{A_{{0}}}^{2}b_{{0}}
\end{gathered}
\end{equation}

\begin{equation}
\begin{gathered}
\label{3.47}b_{{1}}=0, \quad b_{{0}}=-a_{{1}}A_{{0}}
\end{gathered}
\end{equation}

\begin{equation}
\begin{gathered}
\label{3.49}a_{{3}}=-\frac14 a_{{2}}-\frac34\,a_{{0}}-\frac12 a_{{1}}
\end{gathered}
\end{equation}

\begin{equation}
\begin{gathered}
\label{3.50}d_{{1}}=-\frac12 d_{{0}}-\frac32
{A_{{0}}}^{2}a_{{2}}-\frac92{A_{{0}}}^{2}a_{{0 }}-3\,{A_{{0}}}^{2}a_{{1}}
\end{gathered}
\end{equation}

\begin{equation}
\begin{gathered}
\label{3.51}C_{{1}}=-4\,g_{{2}}a_{{2}}-6\,a_{{0}}g_{{2}}
-2\,a_{{1}}g_{{2}}-\\
\\-\frac14{A_{{0}}}^{4}a_{{2}}-\frac34{A_{{0}}}^{4}a_{{0}}
-\frac12{A_{{0}}}^{4}a_{{1}}-\frac12{A_{{0}}}^{2}d_{{0}}
\end{gathered}
\end{equation}

\begin{equation}
\begin{gathered}
\label{3.52}a_{{2}}=-\frac1{6}\,{\frac
{-6\,a_{{1}}g_{{3}}+18\,a_{{0}}g_{{3}}+g_{{2}}d _{{0}}}{g_{{3}}}},
\quad d_{{0}}=\frac {a_{{0}}g_{{2}}^{2}}{g_ {{3}}}
\end{gathered}
\end{equation}

\begin{equation}
\begin{gathered}
\label{3.54}a_{{1}}=\,{\frac {a_{{0}} \left(
{g_{{2}}}^{3}+108\,{g_{{3}}}^{2}
 \right) }{24 g_{{3}}^{2}}}
\end{gathered}
\end{equation}

We also have two relations for constants $g_{{3}}$

\begin{equation}
\begin{gathered}
\label{3.55}g_{{3}}=\pm \frac16\,{g_{{2}}}^{3/2}
\end{gathered}
\end{equation}

As a result we have the following equations

\begin{equation}
\begin{gathered}
\label{3.57}y_{{{\it zzz}}}+ \left( 6\,y-6\,A_{{0}} \right)
y_{{{\it zz}}}-3{y_{{z}
}}^{2}-3\,{y}^{4}+12\,A_{{0}}{y}^{3}+\\
\\
+ \left( -18\,{A_{{0}}}^{2}\mp3\, \sqrt {g_{{2}}} \right) {y}^{2}+
\left(\pm
6\,A_{{0}}\sqrt {g_{{ 2}}}+12\,{A_{{0}}}^{3} \right) y \mp\\
\\
\mp3\,{A_{{0}}}^{2}\sqrt {g_{{2 }}}-3\,{A_{{0}}}^{4}\pm6\,\sqrt
{g_{{2}}}y_{{z}}+6\,g _{{2}}=0
\end{gathered}
\end{equation}
with exact solutions

\begin{equation}
\begin{gathered}
\label{3.58}y \left( z \right) =A_{{0}}+{\frac {R_{{z}}}{R}}
\end{gathered}
\end{equation}

\emph{Fourth order ODEs.} Now let us find nonlinear fourth order
ODEs with exact solutions of the first degree singularity. We have
the following nonlinear fourth order ODE of the general form

\begin{equation}
\begin{gathered}
\label{3.60}a_{{0}}y_{{{\it zzzz}}}+a_{{1}}yy_{{{\it zzz}}}+a_{{2}}{y}^{2}y_{{{ \it
zz}}}+a_{{3}}{y}^{3}y_{{z}}+a_{{4}}{y}^{5}+a_{{5}}y{y_{{z}}}^{2}+a _{{6}}y_{{z}}y_{{{\it
zz}}}+\\
\\
+b_{{0}}y_{{{\it zzz}}}+b_{{1}}yy_{{{\it zz
}}}+b_{{2}}{y_{{z}}}^{2}+b_{{3}}{y}^{4}+d_{{0}}y_{{z}}+d_{{1}}yy_{{z}}
+d_{{2}}{y}^{3}+\\
\\
+h_{{0}}y_{{z}}+h_{{1}}{y}^{2}-C_{{0}}y+C_{{1}}=0
\end{gathered}
\end{equation}

Assuming $g_{{2}}=m^2$ and substituting \eqref{3.31} at
$A_{{0}}=0$ and $A_{{1}}=1$ into \eqref{3.60} we have coefficients
in the form

\begin{equation}
\begin{gathered}
\label{3.62}a_{{6}}=-6\,a_{{0}}-3\,a_{{1}}-4\,a_{{4}}-2\,a_{{2}}-2\,a_{{3}}-a_{{5} }
\end{gathered}
\end{equation}

\begin{equation}
\begin{gathered}
\label{3.63}d_{{2}}=-\frac12 d_{{1}}-\frac12 d_{{0}}
\end{gathered}
\end{equation}

\begin{equation}
\begin{gathered}
\label{3.64}C_{{0}}=-4\,a_{{2}}{m}^{2}-8\,a_{{4}}{m}^{2}+2\,a_{{5}}{m}^{2}
\end{gathered}
\end{equation}

\begin{equation}
\begin{gathered}
\label{3.65}a_{{5}}=a_{{2}}-a_{{3}}-3\,a_{{1}}-\frac16{\frac
{{m}^{2}d_{{1}}}{g_{{3} }}},
\end{gathered}
\end{equation}

\begin{equation}
\begin{gathered}
\label{3.66}a_{{4}}=-\frac32\,a_{{0}}+\frac34\,a_{{1}}-\frac34\,a_{{2}}+\frac14\,a_{{3}}+\frac34\,{
\frac {d_{{0}}g_{{3}}}{{m}^{4}}}-\frac34\,{\frac
{d_{{1}}g_{{3}}}{{m}^{4}} }+\frac{1}{24}\,{\frac
{d_{{1}}{m}^{2}}{g_{{3}}}}
\end{gathered}
\end{equation}

\begin{equation}
\begin{gathered}
\label{3.67}a_{{3}}=-6\,{\frac
{d_{{0}}g_{{3}}}{{m}^{4}}}+\frac{1}{12}\,{\frac {d_{{1}}{m}
^{2}}{g_{{3}}}}+3\,a_{{1}}-15\,a_{{0}}+6\,{\frac
{d_{{1}}g_{{3}}}{{m}^ {4}}}-2\,a_{{2}}
\end{gathered}
\end{equation}

\begin{equation}
\begin{gathered}
\label{3.68}a_{{2}}=-5\,a_{{0}}+2\,a_{{1}}+\frac{1}{36}\,{\frac
{d_{{1}}{m}^{2}}{g_{{3}}}} -3\,{\frac
{d_{{0}}g_{{3}}}{{m}^{4}}}+3\,{\frac {d_{{1}}g_{{3}}}{{m}^{ 4}}}
\end{gathered}
\end{equation}

\begin{equation}
\begin{gathered}
\label{3.69}b_{{3}}=-\frac34b_{{0}}-\frac14b_{{2}}-\frac12b_{{1}}, \quad
h_{{1}}=-\frac12h_{{0}}
\end{gathered}
\end{equation}

\begin{equation}
\begin{gathered}
\label{3.71}C_{{1}}=-6\,b_{{0}}{m}^{2}-4\,b_{{2}}{m}^{2}
\end{gathered}
\end{equation}

\begin{equation}
\begin{gathered}
\label{3.72}b_{{2}}=b_{{1}}-3\,b_{{0}}-\frac16{\frac {h_{{0}}{m}^{2}}{g_{{3}}}}
\end{gathered}
\end{equation}

\begin{equation}
\begin{gathered}
\label{3.73}h_{{0}}=\frac{b_{{0}}m^4}{g_{{3}}}
\end{gathered}
\end{equation}

\begin{equation}
\begin{gathered}
\label{3.74}b_{{0}}=\frac1{16}{\frac
{h_{{0}}g_{{3}}}{{m}^{4}}}+\frac1{36}{\frac {h_{{0}}{
m}^{2}}{g_{{3}}}}
\end{gathered}
\end{equation}

\begin{equation}
\begin{gathered}
\label{3.75}g_{{3}}^{(1,2)}=\pm \frac16{m}^{3}
\end{gathered}
\end{equation}

As a result we have equation in the form at $g_{{3}}=\frac16 m^3$

\begin{equation}
\begin{gathered}
\label{3.77}a_{{0}}y_{{{\it zzzz}}}+\frac16\,\left
(6\,a_{{1}}y+b_{{1}}\right )y_{{{ \it zzz}}}+  \left
(10\,a_{{0}}-a_{{1}}+\frac1{2}\,{\frac {d_{{0}}}{m}}-\frac16\,
{\frac {d_{{1}}}{m}}\right )y_{{z}}y_{{{\it zz}}}
+\\
\\
+\left(d_{{0}}+b_{{1 }}y\right )y_{{{\it zz}}}+\left
(2\,a_{{1}}-\frac23\,{\frac {d_{{1}}}{m}}+ \frac12\,{\frac
{d_{{0}}}{m}}-5\,a_{{0}}\right ){y}^{2}y_{{{\it zz}}}+\\
\\
+ \left (\frac12\,{\frac {d_{{1}}}{m}}+\frac12\,{\frac
{d_{{0}}}{m}}+b_{{2}} \right )y{y_{{z}}}^{2}+\left
(a_{{0}}-\frac12 \,{\frac {d_{{0}}}{m}}-a_{{1}}+\frac13\,{\frac
{d_{{1}}}{m}} \right ){y}^{5}-\\
\\
-\left (\frac14\,b_{{2}}+\frac58\,b_{{1}}\right ){y}^{4}- \left
(\frac12\,d_{{1}}+\frac12\,d_{{0}}\right
){y}^{3}+\frac12\,b_{{1}}m{y}^{2}+ \\
\\
+ \left (d_{{1}}y-5\,{y}^{3}a_{{0}}-{y}^{3}a_{{1}
}-\frac16\,{\frac{{y}^{3}d_{{1}}}{m}}-b_{{1}}m\right )y_{{z}}-\\
\\
-\left(md_{{1}}+12\,{m}^{2}a_{{0}}-3\,md_{{0}}\right )y-4\,b_{{2}}{m}
^{2}-{m}^{2}b_{{1}}=0
\end{gathered}
\end{equation}
with exact solution in the form

\begin{equation}
\begin{gathered}
\label{3.78}y \left( z \right) ={\frac {R_{{z}}}{R}}
\end{gathered}
\end{equation}

Where $R(z)$ is the solution of equation in the form

\begin{equation}
\label{12.1}R_z^2 -4 R^3 + m^2 R+\frac16 m^3 =0
\end{equation}

Solution of the equation \eqref{3.77} is expressed via the
Weierstrass function and has the first degree singularity. At the
present we do not know any application of equation \eqref{3.77}.

\section{Nonlinear ODEs with exact solutions of the second degree singularity}

Let us find nonlinear ODEs with exact solutions of the second
degree singularity expressed via the Weierstrass function

\begin{equation}
\begin{gathered}
\label{4.1}y(z)=A_{{0}}+A_{{1}}{\frac {R_{{z}}}{R}}+A_{{2}}R
\end{gathered}
\end{equation}

We do not consider the second order nonlinear ODEs with exact solutions \eqref{4.1}
because this equation is the exactly solvable equation. We start to find the third order
nonlinear ODEs with exact solutions of the second degree singularity expressed via the
Weierstrass function.

\emph{Third order ODEs}. Let us write the general form of the nonlinear third ODEs with
 solution of the second order singularity. It takes form

\begin{equation}
\begin{gathered}
\label{4.2}a_{{0}}y_{{{\it zzz}}}+a_{{1}}yy_{{z}}+b_{{0}}y_{{{\it zz}}}+b_{{1}}{y
}^{2}+d_{{0}}y_{{z}}-C_{{0}}y+C_{{1}}=0
\end{gathered}
\end{equation}

Substituting \eqref{4.1} into equation \eqref{4.2} we obtain the
following parameters

\begin{equation}
\begin{gathered}
\label{4.3}{a_{1}}=-12\,a_{{0}}, \quad d_{{0}}=12\,a_{{0}}A_{{0}},\quad b_1=\frac
{C_0}{2A_0}
\end{gathered}
\end{equation}

\begin{equation}
\begin{gathered}
\label{4.4}A_{{1}}=0, \quad A_{{0}}=-{\frac {C_{{0}}}{12b_{{0}}}}
\end{gathered}
\end{equation}

\begin{equation}
\begin{gathered}
\label{4.5}g_{{2}}={\frac {{C_{{0}}}^2+24C_{{1}}b_{{0}}}{12b_{{0}}^2}}
\end{gathered}
\end{equation}

As a result we have equation in the form

\begin{equation}
\begin{gathered}
\label{4.6}a_{{0}}y_{{{\it zzz}}}-12a_{{0}}yy_{{z}}+b_{{0}}y_{{{\it zz}}}-{\frac
{a_{{0}}C_{{0}}}{b_{{0}}}}{y_{{z}} }-6b_{{0}}{y^2}-C_{{0}}y+C_{{1}}=0
\end{gathered}
\end{equation}
with exact solutions

\begin{equation}
\begin{gathered}
\label{4.7}y \left( z \right) =-{\frac {C_{{0}}}{12b_{{0}}}}+R \left( z \right)
\end{gathered}
\end{equation}

Equation \eqref{4.6} can be found from the enough popular
nonlinear evolution equation

\begin{equation}
\begin{gathered}
\label{4.8}u_t + \lambda_1 uu_x + \lambda_2 (uu_x)_x + \lambda_3 u_{xx} + \lambda_4
u_{xxx} +\lambda_5 u_{xxxx}=0
\end{gathered}
\end{equation}

Nonlinear evolution equation \eqref{4.8} was used at the
description of nonlinear wave processes and studied in works
\cite{25,26,27,28}. Exact solutions of this equation were obtained
in works \cite{29,30}. They also rediscovered later.

\emph{Fourth order ODEs.} Now let us find the nonlinear fourth
order ODEs with exact solutions of the second degree singularity
expressed via the Weierstrass function. The general case of this
nonlinear ODE can be presented in the form

\begin{equation}
\begin{gathered}
\label{4.14}a_{{0}}y_{{{\it zzzz}}}+a_{{1}}yy_{{{\it zz}}}+a_{{2}}{y_{{z}}}^{2}+a_
{{3}}{y}^{3}+b_{{0}}y_{{{\it zzz}}}+b_{{1}}yy_{{z}}+\\
\\
+d_{{0}}y_{{{\it zz }}}+d_{{1}}{y}^{2}+h_{{0}}y_{{z}}-g_{{2}}C_{{0}}y+C_{{1}}=0
\end{gathered}
\end{equation}

We assume that exact solutions of equation \eqref{4.14} can be found by the formula
\eqref{4.1}.

Without loss of generality let us assume in \eqref{4.1} $A_{{0}}=0$ and $A_{{2}}=1$. We
obtain the following values parameters for
 equation \eqref{4.14}

\begin{equation}
\begin{gathered}
\label{4.17}b_{{1}}=-12\,b_{{0}}, \quad h_{{0}}=0, \quad A_{{1}}=0
\end{gathered}
\end{equation}

\begin{equation}
\begin{gathered}
\label{4.19}a_{{3}}=-120\,a_{{0}}-6\,a_{{1}}-4\,a_{{2}}, \quad d_{{1}}=-6\,d_{{0}}
\end{gathered}
\end{equation}

\begin{equation}
\begin{gathered}
\label{4.21}a_{{2}}=-18a_{{0}}-\frac12 a_{{1}}-C_{{0}}
\end{gathered}
\end{equation}

\begin{equation}
\begin{gathered}
\label{4.22}C_{{1}}=\frac12\,d_{
{0}}g_{{2}}-\frac12\,g_{{3}}a_{{1}}-6\,a_{{0}}g_{{3}}-g_{{3}}C_{{0}}
\end{gathered}
\end{equation}

We get nonlinear ODEs in the form

\begin{equation}
\begin{gathered}
\label{4.23}a_{{0}}y_{{{\it zzzz}}}+b_{{0}}y_{{{\it zzz}}}+\left
(a_{{1}}y+d_{{0}} \right )y_{{{\it zz}}}-\left
(18\,a_{{0}}+\frac12\,a_{{1}}+C_{{0}}\right
){y_{{z}}}^{2}-\\
\\
-12\,b_{{0}}yy_{{z}}+\left (4\,C_{{0}}-48\,a_{{0}}-4\,a_
{{1}}\right ){y}^{3}-6\,d_{{0}}{y}^{2}-g_{{2}}C_{{0}}y+\\
\\+\frac12\,d_{{0}}g_{{2}}-\frac12\,g_{{3}}a_{{1}}-
6\,a_{{0}}g_{{3}}-g_{{3}}C_{{0}}=0
\end{gathered}
\end{equation}

Exact solutions of equation \eqref{4.23} is found by the formula

\begin{equation}
\begin{gathered}
\label{4.16}y \left( z \right) =R \left( z \right)
\end{gathered}
\end{equation}

Assuming $a_{{0}}=1$, $b_{{0}}=0$, $d_{{0}}=0$ in \eqref{4.23} we
have equation

\begin{equation}
\begin{gathered}
\label{4.16d}y_{{{\it zzzz}}}+a_{{1}}yy_{{{\it zz}}}-\left
(18\,+\frac12 \,a_{{1}}+C_{{0}}\right ){y_{{z}}}^{2}+\\
\\+
4\left (C_{{0}}-12\,-a_{{1}}\right
){y}^{3}-g_{{2}}C_{{0}}y-6\,g_{{3}}-\frac12\,g_{{3}
}a_{{1}}-g_{{3}}C_{{0}}=0
\end{gathered}
\end{equation}

One can note that the last equation has two parameters in leading
members. However let us show that this equation contains a lot of
important nonlinear integrable and nonintegrable differential
equations.

Assuming $a_{{1}}=-30$, $C_{{0}}=-3$ and $y(z)\rightarrow -y(z)$ in equation
\eqref{4.16d} we have equation in the form

\begin{equation}
\begin{gathered}
\label{4.22a}y_{{{\it zzzz}}}+30\,yy_{{{\it
zz}}}+60\,{y}^{3}+3\,g_{{2}}y-12\,g_{{3 }}=0
\end{gathered}
\end{equation}

Equation \eqref{4.22a} is exactly solvable equation. This equation can be obtained by the
travelling wave \eqref{1.2} from the Caudrey -- Dodd -- Gibbon equation \cite{30a,31,32}

\begin{equation}
\begin{gathered}
\label{4.22d}u_{{t}}+\frac {\partial}{\partial
x}\left(u_{{xxxx}}+30uu_{{xx}}+60u^3\right)=0
\end{gathered}
\end{equation}

Equation \eqref{4.22d} is integrable equation by the inverse scattering transform.
However we can see that this equation is found in our class of nonlinear differential
equations.

Assuming $a_{{1}}=-20$, $C_{{0}}=2$ and $y(z)\rightarrow -y(z)$ in equation \eqref{4.16d}
we have equation in the form

\begin{equation}
\begin{gathered}
\label{4.23a}y_{{{\it zzzz}}}+20\,yy_{{{\it
zz}}}+10\,{y_{{z}}}^{2}+40\,{y}^{3}-2\, g_{{2}}y-2\,g_{{3}}=0
\end{gathered}
\end{equation}

Using the travelling wave we can see that equation \eqref{4.23a}
is obtained from the Korteveg - de Vries equation of the fifth
order \cite{31,32}

\begin{equation}
\begin{gathered}
\label{4.22c}u_{{t}}+\frac{\partial}{\partial
x}\left(u_{{xxxx}}+20uu_{{xx}}+10u_{{x}}^2+40u^3\right)=0
\end{gathered}
\end{equation}

We have obtained that special solutions of the fifth order Korteveg - de Vries  equation
can be found by the formula \eqref{4.16}.

Assuming $a_{{1}}=-15$, $C_{{0}}=\frac 34$ and $y(z)\rightarrow
-\frac23 y(z)$ in \eqref{4.16d} we have equation with exact
solution \eqref{4.16} in the form

\begin{equation}
\begin{gathered}
\label{4.233}y_{{zzzz}}+10\,yy_{{zz}}+\frac{15}{2}y_{{z}}^2+{\frac
{20}{3}}y^{3}-\frac34\,g_{{2}}y-{\frac {9}{8}}\,g_{{3}}=0
\end{gathered}
\end{equation}

Equation \eqref{4.233} is exactly solvable equation too. Using the travelling wave this
one can be obtained from the Kaup - Kupershmidt equation \cite {31,31a,32}

\begin{equation}
\begin{gathered}
\label{4.222}u_{{t}}+\frac{\partial}{\partial x}
{\left(u_{{xxxx}}+10uu_{{xx}}+\frac{15}2u_{{x}}^2+
\frac{20}3u^3\right)}=0
\end{gathered}
\end{equation}

Assuming $a_{{1}}=-15$, $C_{{0}}=\frac 34$ and $y(z)\rightarrow -
y(z)$ in \eqref{4.16d} we have equation

\begin{equation}
\begin{gathered}
\label{4.234}y_{{{\it zzzz}}}-15\,yy_{{{\it zz}}}+{\frac
{45}{4}}\,{y_{{z}}}^{2}+15
\,{y}^{3}-\frac34\,g_{{2}}y-\frac34\,g_{{3}}=0
\end{gathered}
\end{equation}

Equation \eqref{4.234} is also exactly solvable equation \cite {32a}. This one has exact
solution \eqref{4.16} as well. This equation can be found from the Schwarz -- Kaup--
Kuperschmidt equation of the fifth order, that is the singular manifold equation for the
Kaup -- Kuperschmidt equation. The Cauchy problems for these equations can be solved by
inverse scattering transform.

Assuming $a_{{1}}=-12$, $C_{{0}}=0$ and $y(z)\rightarrow - y(z)$
in \eqref{4.16d} we have equation

\begin{equation}
\begin{gathered}
\label{4.235}y_{{{\it zzzz}}}+12\,yy_{{{\it
zz}}}+12\,{y_{{z}}}^{2}=0
\end{gathered}
\end{equation}

Equation \eqref{4.235} is exactly solvable equation again \cite {32a}. The general
solution of this equation is expressed via the first Painleve transcendent. However this
equation has special solution expressed by the formula \eqref{4.16} as well.

We have interesting result that is a number of exactly solvable equations are found in
our class of differential equations. Much more than that these nonlinear ODEs have the
similar special solutions expressed by the formula \eqref{4.16} via the Weierstrass
function. We are going to use this observation in future work to look for exactly
solvable nonlinear ODEs of higher order.

Assuming $a_{{1}}=0$ and $C_{{0}}=-18a_{{0}}$ in \eqref{4.23} we
have equation

\begin{equation}
\begin{gathered}
\label{4.24}a_{{0}}y_{{{\it zzzz}}}+b_{{0}}y_{{{\it zzz}}}+d_{{0}}y_{{{\it zz}}}-
12\,b_{{0}}yy_{{z}}-120\,{y}^{3}a_{{0}}-\\
\\
-6\,d_{{0}}{y}^{2}+18\,g_{{2}}a _{{0}}y+C_{{1}}=0
\end{gathered}
\end{equation}

At $b_{{0}}=0$ we have nonlinear ODEs

\begin{equation}
\begin{gathered}
\label{4.24a}a_{{0}}y_{{{\it zzzz}}}+d_{{0}}y_{{{\it zz}}}-
120a_{{0}}\,{y}^{3}-6\,d_{{0}}{y}^{2}+48\,g_{{2}}a
_{{0}}y+C_{{1}}=0
\end{gathered}
\end{equation}

Last equation corresponds to the nonlinear evolution equation that
is used at the description of the nonlinear dispersive waves
\cite{33,34,35,36}

\begin{equation}
\begin{gathered}
\label{4.19}u_t +\alpha uu_x + \beta u^2 u_x +\gamma u_{xxx}
+\delta u_{xxxxx}=0
\end{gathered}
\end{equation}

We have equation \eqref{4.24a} if we look for exact solution of
equation \eqref{4.19} in the form of the travelling wave
\eqref{1.2}.

In the case $d_{{0}}=0$ from \eqref{4.24} we obtain the nonlinear
ODEs in the form

\begin{equation}
\begin{gathered}
\label{4.25}a_{{0}}y_{{{\it
zzzz}}}-120a_{{0}}\,{y}^{3}+48a_{{0}}\,g_{{2}}y+C_{{1} }=0
\end{gathered}
\end{equation}

Equation \eqref{4.25} are used at the description of nonlinear
dispersive waves as well. Corresponding nonlinear evolution
equation can be met at the the description of the nonlinear waves
and takes the form \cite{37,38}

\begin{equation}
\begin{gathered}
\label{4.21}u_t + \beta u^2 u_x +\delta u_{xxxxx}=0
\end{gathered}
\end{equation}

Exact solutions of equations \eqref{4.24},\eqref{4.24a} and
\eqref{4.25} are found by the formula \eqref{4.16}.

\section{Nonlinear ODEs with the third degree singularity
solution}

In this section we are going to find nonlinear ODEs that have
exact solutions of the third degree singularity and are expressed
via the Weierstrass function. These solutions can be presented by
the formula

\begin{equation}
\begin{gathered}
\label{5.1}y \left( z \right) =A_{{0}}+A_{{1}}{\frac {
R_{{z}}}{R}}+A_{{2}}R + A_{{3}}R_{{z}}
\end{gathered}
\end{equation}

We can not suggest any nonlinear ODEs of the second order with
exact solutions \eqref{5.1} because we can not have the polynomial
class of this equation. Let us start to consider the third
nonlinear ODE.

\emph{Third order ODE.} We have the general case of the nonlinear
third order ODE in the form

\begin{equation}
\begin{gathered}
\label{5.2}a_{{0}}y_{{zzz}}  +a_{{1}} y^{2}
 +b_{{0}}y_{{zz}} +d_{{0}}y _{{z}}-C_{{0}}y
  +C_{{1}}=0
\end{gathered}
\end{equation}

Assuming $A_{{3}}=1$ and $g_{{2}}={\frac{n^2}{12}}$ we have
following relations for the parameters

\begin{equation}
\begin{gathered}
\label{5.3}d_{{0}}=-2a_{{1}}A_{{1}}+6{\frac
{b_{{0}}A_{{1}}}{A_{{2}}}}+{\frac {C_{{0}}}{A_{{2}}}}-2\,{\frac
{a_{{1}}A_{{0}}}{A_{{2}}}}
\end{gathered}
\end{equation}

\begin{equation}
\begin{gathered}
\label{5.4}A_{{1}}=0, \quad b_{{0}}=-a_{{0}}A_{{2}}-{\frac{1}{6}}a_{{1}}A_{{2}}
\end{gathered}
\end{equation}

\begin{equation}
\begin{gathered}
\label{5.5}a_{{1}}=-30\,a_{{0}},\quad C_{{0}}=-a_{{0}} \left( 60\,A_{{0}}-{A_{{2}}}^{3}
\right)
\end{gathered}
\end{equation}

We obtain four values for constant $A_{{2}}$

\begin{equation}
\begin{gathered}
\label{5.6}{A_{{2}}^{{(1,2)}}=\pm{n}}, \quad {A_{{2}}^{{(3,4)}}=\pm{i n}}
\end{gathered}
\end{equation}
and four values for constant $C_{{1}}$

\begin{equation}
\begin{gathered}
\label{5.7}C_{{1}}^{{(1,2)}}=-{\frac{a_{{0}}}{24}}
\left(432g_{{3}}+720A_{{0}}^2\mp24A_{{0}}n^3-5n^6
\right)
\end{gathered}
\end{equation}

\begin{equation}
\begin{gathered}
\label{5.8}C_{{1}}^{{(3,4)}}=i{\frac{a_{{0}}}{24}}\left(432ig_{{3}}+720iA_{{0}}^2\mp24A_{{0}}n^3+5in^6
\right)
\end{gathered}
\end{equation}

We have four nonlinear ODEs

\begin{equation}
\begin{gathered}
\label{5.9}a_{{0}}y_{{zzz}}-30a_{{0}}{y}^{2}\pm4a_{{0}}ny_{{zz}}+a_{{0}}n^{2}y_{{z}}{\pm}a_{{0}}n^{3}y
+60a_{{0}}A_{{0}}y+C_{{1}}=0
\end{gathered}
\end{equation}

\begin{equation}
\begin{gathered}
\label{5.11}a_{{0}}y_{{zzz}}-30a_{{0}}{y}^{2}\pm4ia_{{0}}ny_{{zz}}-a_{{0}}n^{2}y_{{z}}{\mp}i{a_{{0}}}n^{3}y
+60a_{{0}}A_{{0}}y+C_{{1}}=0
\end{gathered}
\end{equation}
with exact solutions

\begin{equation}
\begin{gathered}
\label{5.61d}y \left( z \right) =A_{{0}}\pm{nR} + R_{{z}}
\end{gathered}
\end{equation}

\begin{equation}
\begin{gathered}
\label{5.61e}y \left( z \right) =A_{{0}}\pm{i}{nR} + R_{{z}}
\end{gathered}
\end{equation}

Equation \eqref{5.9} can be found from nonlinear evolution
equation

\begin{equation}
\begin{gathered}
\label{5.10}u_t + \alpha uu_x+\beta u_{{xx}}+\gamma
u_{{xxx}}+\delta u_{xxxx} =0
\end{gathered}
\end{equation}

This is the Kuramoto -- Sivashinsky equation \cite{39,40,40a}.
Equation \eqref{5.10} is used at the description of turbulence
processes \cite{40a,41,42}. Solution of this equation was found in
works \cite{10,12,14,15}.

 From equation \eqref{5.9} and \eqref{5.11} one can see there is the special solution of
 equation \eqref{5.10} at $\beta=0$ and $\gamma=0$. This equation reminds the Burgers
 equation and takes the form

\begin{equation}
\begin{gathered}
\label{5.10a}u_t + \alpha uu_x+\delta u_{xxxx} =0
\end{gathered}
\end{equation}

Exact solutions of equation \eqref{5.10a} is found by the formula

\begin{equation}
\begin{gathered}
\label{5.71d}y \left( z \right) =A_{{0}} + R_{{z}}
\end{gathered}
\end{equation}
if we look for solution of equation \eqref{5.10a} in the form of the travelling wave.
This equation takes the form

\begin{equation}
\begin{gathered}
\label{5.71e}\delta y_{{zzzz}}+\frac{\alpha}2
y^2-C_{{0}}y+\frac{C_{{0}}^2+2160\delta^2g_{{3}}}{2\alpha}=0
\end{gathered}
\end{equation}

There is the rational solution of equation \eqref{5.71e} in the
form

\begin{equation}
\begin{gathered}
\label{5.72d}y \left( z \right) =\frac{C_{{0}}}{\alpha} +
\frac{120\delta}{\alpha(z-z_{{0}})^3}
\end{gathered}
\end{equation}

In this case equation \eqref{5.71e} can be written in the simple
form

\begin{equation}
\begin{gathered}
\label{5.73d}\delta y_{{zzzz}}+\frac{\alpha}2
y^2-C_{{0}}y+\frac{C_{{0}}^2}{2\alpha}=0
\end{gathered}
\end{equation}

We found nonlinear ODEs that correspond to the Kuramoto - Sivashinsky equation. These
equations are very popular at the description of the turbulence processes.

\emph{Fourth order ODEs.} Let us find the nonlinear ODEs of fourth
order with exact solutions \eqref{5.1}. We can use the following
general form of the fourth order

\begin{equation}
\begin{gathered}
\label{5.13}a_{{0}}y_{{{\it zzzz}}}+a_{{1}}yy_{{z}}+b_{{0}}y_{{{\it zzz}}}+b_{{1}}
{y}^{2}+d_{{0}}y_{{{\it zz}}}+h_{{0}}y_{{z}}-g_{{2}}C_{{0}}y+C_{{1}}=0
\end{gathered}
\end{equation}

Assuming $A_{{2}}=-60$ and substituting \eqref{5.1} at
$g_{{2}}=6A_{{2}}^4$ into equation \eqref{5.13} we obtain at
$A_{{1}}=0$ relations for the parameters

\begin{equation}
\begin{gathered}
\label{5.15}a_{{1}}=a_{{0}}
\end{gathered}
\end{equation}

\begin{equation}
\begin{gathered}
\label{5.16}d_{{0}}=-{\frac{1}{6}}b_{{1}}A_{{2}}+{\frac
{1}{60}}\,b_{{0}}A_{{2}}+{\frac {1 }{720}}\,a_{{0}}{A_{{2}}}^{2}
\end{gathered}
\end{equation}

\begin{equation}
\begin{gathered}
\label{5.17}h_{{0}}=4320\,a_{{0}}{A_{{2}}}^{3}-360\,{A_{{2}}}^{3}C_{{0}}
\end{gathered}
\end{equation}

\begin{equation}
\begin{gathered}
\label{5.18}b_{{1}}={\frac{1}{2}}b_{{0}}+{\frac{1}{30}}a_{{0}}A_{{2}}
\end{gathered}
\end{equation}

\begin{equation}
\begin{gathered}
\label{5.19}b_{{0}}={\frac{1}{12}}\,A_{{2}} \left(
186623999\,a_{{0}}-15552000\,C_{{0}}
 \right)
\end{gathered}
\end{equation}

\begin{equation}
\begin{gathered}
\label{5.20}C_{{0}}=12a_{{0}}
\end{gathered}
\end{equation}

\begin{equation}
\begin{gathered}
\label{5.21}C_{{1}}=-18\,a_{{0}}A_{{2}}g_{{3}}+{ \frac
{1}{240}}\,{A_{{2}}}^{7}a_{{0}}
\end{gathered}
\end{equation}

Taking these values for the parameters into account we have
nonlinear ODE in the form

\begin{equation}
\begin{gathered}
\label{5.22}a_{{0}}y_{{{\it zzzz}}}-{\frac {1}{12}}\,A_{{2}}a_{{0}
}y_{{{\it zzz}}}+{\frac {1}{720}}\,a_{{0}}{A_{{2}}}^{ 2}y_{{{\it
zz}}}+ a_{{0}}yy_{{z}}-{\frac
{1}{120}}\,a_{{0}}A _{{2}}{y}^{2}-\\
\\
-{72}\,{A_{{2}}}^{4}a_{{0}}y-{18}\,a_{{0}}A_{{2}}g_{{3}}+{\frac {
1}{240}}\,{A_{{2}}}^{7}a_{{0}}=0
\end{gathered}
\end{equation}

Equation \eqref{5.22} can be found from the nonlinear evolution equation

\begin{equation}
\begin{gathered}
\label{5.23}u_t + \alpha (uu_x)_x + \beta uu_x + \gamma u_{xx} + \delta u_{xxx}
+\varepsilon u_{xxxx} +u_{xxxxx}=0
\end{gathered}
\end{equation}
if we look for exact solution in the form of the travelling wave \eqref{1.2}.

Solution of this equation is expressed by the formula

\begin{equation}
\begin{gathered}
\label{5.24}y \left( z \right) =A_{{2}}R-60\,R_{{z}}
\end{gathered}
\end{equation}

At $A_{{2}}=0$ from \eqref{5.22} we obtain the nonlinear ODE that
corresponds to the partial case \eqref{5.10a} of the Kuramoto -
Sivashinsky equation again.

\section{Nonlinear ODEs with exact solutions of the fourth order singularity}

Let us find nonlinear ordinary differential equations which have
exact solutions of the fourth order singularity. These solutions
can be presented by the formula

\begin{equation}
\begin{gathered}
\label{6.1}y \left( z \right) =A_{{0}}+{\frac {A_{{1}}R_{{z}}}{R}}+A_{{2}}R+A_{{3
}}R_{{z}}+A_{{4}}{R}^{2}
\end{gathered}
\end{equation}
where $R(z)$ satisfies the Weierstrass elliptic equation again.

We can not suggest nonlinear ODEs of the second and third order of the polynomial form
with solution \eqref{6.1}. However we can consider the nonlinear fourth order ODE that
takes the form.

\begin{equation}
\begin{gathered}
\label{6.2}a_{{0}}y_{{{\it zzzz}}}+a_{{1}}{y}^{2}+d_{{0}}y_{{{\it zz}}}+h_{{0}}y_
{{z}}-g_{{2}}C_{{0}}y+C_{{1}}=0
\end{gathered}
\end{equation}

For calculations it is convenient to use  $A_1 =0,\,\,\, A_3 =0$ and $A_4=-180$.

\begin{equation}
\begin{gathered}
\label{6.2a}y \left( z \right) =A_{{0}}+A_{{2}}R-180\,{R}^{2}
\end{gathered}
\end{equation}

Substituting \eqref{6.1} into equation \eqref{6.2} we have

\begin{equation}
\begin{gathered}
\label{6.3b}h_0 =0,\quad a_{{1}}=\frac{14}3\,a_{{0}}, \quad d_{{0}}=-{\frac
{13}{30}}\,a_{{0}}A_{{2}}
\end{gathered}
\end{equation}

\begin{equation}
\begin{gathered}
\label{6.4}A_{{0}}={\frac {3}{28}}\,{\frac {g_{{2}}C_{{0}}}{a_{{0}}}}+{\frac {31}
{25200}}\,{A_{{2}}}^{2}+18\,g_{{2}}
\end{gathered}
\end{equation}

\begin{equation}
\begin{gathered}
\label{6.5}g_{{3}}=-{\frac {1}{58320000}}\,A_{{2}} \left( 31\,{A_{{2}}}^{2}-
226800\,g_{{2}} \right)
\end{gathered}
\end{equation}

\begin{equation}
\begin{gathered}
\label{6.6}C_{{1}}=-{\frac {127}{300}}\,a_{{0}}{A_{{2}}}^{2}g_{{2}}-{\frac {961}{
136080000}}\,a_{{0}}{A_{{2}}}^{4}+{\frac {3}{56}}\,{\frac {{g_{{2}}}^{
2}{C_{{0}}}^{2}}{a_{{0}}}}-\\
\\
-1242\,a_{{0}}{g_{{2}}}^{2}+168\,a_{{0}}A_{{ 2}}g_{{3}}
\end{gathered}
\end{equation}

In this case we get equation

\begin{equation}
\begin{gathered}
\label{6.10}a_{{0}}y_{{{\it zzzz}}}+\frac{14}3\,a_{{0}}{y}^{2}-{\frac {13}{30}}\,a_{{0}}
A_{{2}}y_{{{\it zz}}}-g_{{2}}C_{{0}}y-{\frac {127}{300}}\,a_{{0}}{A_{{ 2}}}^{2}g_{{2}}-
\\
\\
-{\frac {961}{136080000}}\,a_{{0}}{A_{{2}}}^{4}+{\frac {3}{56}}\,{\frac
{{g_{{2}}}^{2}{C_{{0}}}^{2}}{a_{{0}}}}-1242\,a_{{0}}{
g_{{2}}}^{2}+168\,a_{{0}}A_{{2}}g_{{3}}=0
\end{gathered}
\end{equation}
with exact solution

\begin{equation}
\begin{gathered}
\label{6.11}y \left( z \right) =A_{{0}}+A_{{2}}R-180\,{R}^{2}
\end{gathered}
\end{equation}

Equation \eqref{6.10} corresponds to the nonlinear evolution
equation \cite{33,34,38}

\begin{equation}
\begin{gathered}
\label{6.12}u_t + \alpha uu_x + \beta u_{xxx} =\delta u_{xxxxx}
\end{gathered}
\end{equation}

Exact solutions of equation \eqref{6.10} were found before in
works \cite{10,13} and rediscovered in a number of papers later.

At $A_{{2}}=0$ from \eqref{6.10} we have nonlinear ODE in the form

\begin {equation}
\begin{gathered}
\label{6.13} a_{{0}}y_{{{\it
zzzz}}}+\frac{14}3\,a_{{0}}{y}^{2}-g_{{2}}C_{{0}}y-1242\,a_{
{0}}{g_{{2}}}^{2}+{\frac {3}{56}}\,{\frac
{{g_{{2}}}^{2}{C_{{0}}}^{2}} {a_{{0}}}}=0
\end{gathered}
\end{equation}

This equation can be found from the nonlinear evolution equation
\cite{34,38}

\begin {equation}
\begin{gathered}
\label{6.14}u_{{t}}+\alpha\,uu_{{x}}+\delta\,u_{{{\it xxxxx}}}=0
\end{gathered}
\end{equation}

if we look for exact solutions  of this nonlinear evolution equation in the form of the
travelling wave. Special solution of equations \eqref{6.13} and \eqref{6.14} are found by
the formula \eqref{6.11} at $A_{{2}}=0$.

\section{Nonlinear ODE with exact solution of the fifth degree singularity}

We have not got any possibility to suggest any nonlinear
differential equation of the polynomial class  of the second,
third and fourth order with exact solution of the fifth degree
singularity. We can look for such type of nonlinear fifth order
ODE.

Assume we want to have nonlinear ODE with exact solution of the
fifth degree singularity

\begin{equation}
\begin{gathered}
\label{7.1}y \left( z \right) =A_{{0}}+{\frac {A_{{1}}R_{{z}}}{R}}+A_{{2}}R+A_{{3
}}R_{{z}}+A_{{4}}{R}^{2}+A_{{5}}RR_{{z}}
\end{gathered}
\end{equation}

Simplest case of the nonlinearity for this solution takes the form $y^2$ and we can see
that \eqref{7.1} give singularity of the tenth degree. General case of equation takes the
form

\begin{equation}
\begin{gathered}
\label{7.2}\varepsilon\,y_{{{\it zzzzz}}}+\gamma\,
y_{{zzzz}}+\delta\,y_{{{\it zzz}}}+\chi\,
y_{{zz}}+\beta\,y_{{z}}+\alpha\,{y}^{2} -C_{{0}}y +C_{{1}}=0
\end{gathered}
\end{equation}

This equation can be found from the nonlinear evolution equation
\cite{18,43}

\begin{equation}
\begin{gathered}
\label{7.2a}u_{{t}}+2\alpha\,u u_{{x}}+\beta\,u_{{xx}}+\chi\,
u_{{xxx}}+\delta\,u_{{{\it xxxx}}}+\gamma\,
u_{{xxxxx}}+\varepsilon u_{{{\it xxxxxx}}}=0
\end{gathered}
\end{equation}
if we search exact solutions in the form of the travelling wave \eqref{1.2}.

Consider the simplest case of equation \eqref{7.2a} taking
$\gamma=0$, $\chi=0$ and $\alpha=\frac12$ into account. In this
case from \eqref{7.2} we get equation

\begin{equation}
\begin{gathered}
\label{7.2b}\varepsilon\,y_{{{\it zzzzz}}}+\delta\,y_{{{\it
zzz}}}+\beta\,y_{{z}}+\frac12\,{y}^{2} -C_{{0}}y +C_{{1}}=0
\end{gathered}
\end{equation}

Let us find exact solutions of the nonlinear ODE \eqref{7.2b}. At $A_1=0,\,A_2=0,\,
A_4=0$ we have relations for constants

\begin{equation}
\begin{gathered}
\label{7.3}A_{{5}}=-15120\,\varepsilon, \quad A_{{3}}=-{\frac
{1260}{11}}\,\delta
\end{gathered}
\end{equation}

\begin{equation}
\begin{gathered}
\label{7.4}\beta={\frac {10}{121}}\,{\frac
{{\delta}^{2}}{\varepsilon}}
\end{gathered}
\end{equation}

\begin{equation}
\begin{gathered}
\label{7.5} g_{{3}}=-{\frac {7}{660}}\,{\frac
{\delta\,g_{{2}}}{\varepsilon}}-{\frac {1}{574992}}\,{\frac
{{\delta}^{3}}{{\varepsilon}^{3}}} ,\quad g_{{2}}={\frac
{1}{1452}}\,{\frac {{\delta}^{2}}{{\varepsilon}^{2}}}+{ \frac
{1}{5082}}\,{\frac {{\delta}^{2}\sqrt {21}}{{\varepsilon}^{2}}}
\end{gathered}
\end{equation}

\begin{equation}
\begin{gathered}
\label{7.6}C_{{1}}=-{\frac {10854}{161051}}\,{\frac
{{\delta}^{5}}{{\varepsilon}^{3} }}-{\frac {2484}{161051}}\,{\frac
{{\delta}^{5}\sqrt {21}}{{\varepsilon}^ {3}}}+\frac12{C_{{0}}}^{2}
\end{gathered}
\end{equation}

Substituting \eqref{7.4} and \eqref{7.6} into equation \eqref{7.2} we obtain nonlinear
ODE in the form

\begin{equation}
\begin{gathered}
\label{7.7}\varepsilon\,y_{{{\it zzzzz}}}+\delta\,y_{{{\it
zzz}}}+{\frac {10}{121}}\,{ \frac
{{\delta}^{2}y_{{z}}}{\varepsilon}}+\frac12{y}^{2}-C_{{0}}y +\\
\\
+\frac12{C_{{0}}}^{2}-{\frac {10854}{161051}}\,{\frac
{{\delta}^{5}}{{\varepsilon}^{3} }}-{\frac {2484}{161051}}\,{\frac
{{\delta}^{5}\sqrt {21}}{{\varepsilon}^ {3}}}=0
\end{gathered}
\end{equation}
with exact solution

\begin{equation}
\begin{gathered}
\label{7.8}y \left( z \right) =C_{{0}}-\frac{1260}{11} R_z
\left(\delta +132\varepsilon R\right)
\end{gathered}
\end{equation}

Where $R(z)$ satisfies the Weierstrass elliptic equation

\begin{equation}
\begin{gathered}
\label{7.18} R_{{z}}^{2}-4\,R^{3}+ \left ({\frac
{1}{1452}}\,{\frac {{\delta}^{2}}{{\epsilon}^{2}}}+{ \frac
{1}{5082}}\,{\frac {{\delta}^{2}\sqrt {21}}{{\epsilon}^{2}}}
\right )R-{\frac {7}{660}}\,{\frac {\delta\,g_{{2}}}{\epsilon}}-{
\frac {1}{574992}}\,{\frac {{\delta}^{3}}{{\epsilon}^{3}}}=0
\end{gathered}
\end{equation}

Equation \eqref{7.7} is found from the nonlinear evolution equation

\begin{equation}
\begin{gathered}
\label{7.9}u_t + uu_x + \beta u_{xx} + \delta u_{xxxx} + \varepsilon u_{xxxxxx}=0
\end{gathered}
\end{equation}

In the case $\delta=0$ we obtain $\beta=0$ and have the rational solution of equation
\eqref{7.9}. This solution takes the form

\begin{equation}
\begin{gathered}
\label{7.10}y(z)=C_{{0}} + \frac{30240\varepsilon}{(z+z_{{0}})^5}
\end{gathered}
\end{equation}

Equation \eqref{7.7} in this case takes the form

\begin{equation}
\begin{gathered}
\label{7.10a}\varepsilon\,y_{{{\it
zzzzz}}}+\frac12{y}^{2}-C_{{0}}y +\frac12{C_{{0}}}^{2}=0
\end{gathered}
\end{equation}

Last years equation \eqref{7.9} is used at description of extensive model model chaos
\cite{44,45,46}. We hope that exact solution \eqref{7.8} will be useful at the study of
the turbulence processes where equation \eqref{7.9} is applied.

Rational solutions \eqref{5.72d} and \eqref{7.10} of equations
\eqref{5.73d} and \eqref{7.10a} suggest us how one can find
rational solutions of the generalized differential equations

\begin{equation}
\begin{gathered}
\label{7.12}u_t + \alpha uu_x +  \varepsilon u_{n+1,x}=0,\, \quad
u_{{n+1,x}}=\frac{\partial^{(n+1)} u}{\partial x^{(n+1)}}
\end{gathered}
\end{equation}

Using the travelling wave \eqref{2.1} we have from \eqref{7.12} the nonlinear ODE in the
form

\begin{equation}
\begin{gathered}
\label{7.16}\varepsilon\,y_{{{\it n,z}}}+\frac {\alpha}
{2}{y}^{2}-C_{{0}}y +\frac12{C_{{0}}}^{2}=0, \, \quad
y_{{n,x}}=\frac{d^n y}{d x^n}
\end{gathered}
\end{equation}

Rational solutions of \eqref{7.16} takes the form

\begin{equation}
\begin{gathered}
\label{7.17} y(z)=\frac {C_{{0}}}{\alpha}+\frac
{2\,(-1)^{(n-1)}\,(2n-1)!\,\varepsilon}{\alpha\,
(n-1)!\,(z+z_{{0}})^{n}}
\end{gathered}
\end{equation}

Here $z_{{0}}$ is a arbitrary constant. Exact solution
\eqref{7.17} can be useful at numerical simulation of the
nonlinear problems where equations \eqref{7.12} and \eqref{7.16}
are used.

\newpage

\section{Conclusion}

Let us emphasize in brief the results of this work. We noted that
a lot of nonlinear differential equations have exact solutions
expressed via the Weierstrass function.

This observation provided the idea to find nonlinear differential
equations with special solutions expressed via the Weierstrass
function. We formulated new problem and found the polynomial class
of nonlinear differential equations of the second, third and
fourth order which have exact solutions expressed via the
Weierstrass function. Exact solutions of these equations have
different singularities and are expressed via general solutions of
the Weierstrass elliptic equation.

We also list a number of nonlinear ODEs with exact solutions.
These equations are found from the widely used nonlinear evolution
equations taking the travelling wave into account.

\section {Acknowledgments}

This work was supported by the International Science and Technology Center under Project
No 1379-2. This material is partially based upon work supported by the Russian Foundation
for Basic Research under Grant No 01-01-00693.

\end{document}